\documentclass[showpacs]{revtex4}
\usepackage{graphicx,amsmath,amssymb,amsfonts,bbm,latexsym,color,dcolumn,epsf}

\def\be{\begin{equation}}
\def\ee{\end{equation}}
\def\ba{\begin{eqnarray}}
\def\ea{\end{eqnarray}}

\begin{document}

\title{Running of Newton's constant and non integer powers of the d'Alembertian}
\author{D. L\'opez Nacir \footnote{dnacir@df.uba.ar}}
\author{F. D. Mazzitelli \footnote{fmazzi@df.uba.ar}}
\affiliation{Departamento de F\'\i sica {\it Juan Jos\'e
Giambiagi}, Facultad de Ciencias Exactas y Naturales, UBA, Ciudad
Universitaria, Pabell\' on I, 1428 Buenos Aires, Argentina}

\begin{abstract}
The running of Newton's constant can be taken into account by
considering covariant, non local generalizations of the field
equations of general relativity. These generalizations involve
nonanalytic functions of the d'Alembertian, as
$(-\Box)^{-\alpha}$, with $\alpha$ a non integer number, and
$\ln[-\Box]$. In this paper we define these non local operators in
terms of the usual two point function of a massive field. We
analyze some of their properties, and present specific
calculations in flat and Robertson Walker
spacetimes.
\end{abstract}


\maketitle

\section{Introduction}

In quantum field theory, the coupling constants depend on the
energy scale according to the Renormalization Group Equations
(RGE). This running behavior is crucial to understand perturbative
and nonperturbative aspects of elementary particles and their
interactions.

Over the years, there have been several  attempts to incorporate
the running behavior in astrophysics and cosmology. In the first
case, a scale dependent Newton's constant would explain at least
part of the dark matter. There are proposals in the literature
\cite{gder} where it is assumed  that the gravitational constant
around a compact object depends on the radial coordinate $r$. The
dependence is dictated by the RGE, replacing the energy scale
$\bar\mu$ by $1/r$.

In a cosmological framework, the expansion of the universe makes
any physical length scale to increase with time and naively one
would expect the coupling constants to become time dependent as
the universe evolves. Hopefully, this time dependence would help
to explain the accelerated expansion of the Universe and to shed
light into the dark energy problem. This point of view has  been
considered by a number of authors, and several proposals have been
analyzed in detail. For example, it was assumed  \cite{gdet} that
Newton's and cosmological constants depend on $t$ according to the
RGE through the replacement $\bar\mu \rightarrow 1/t$. The
Einstein equations with these time dependent constants in general
do not satisfy Bianchi identites, and therefore some constraints
are needed for mathematical consistency \cite{Babic}. Related
proposals assume that the correct replacement is obtained with the
Hubble ``constant" $\bar\mu \rightarrow  H$ \cite{gdeH}, the
inverse radii of the event horizon \cite{gdeho}, or with the scale
factor $\bar\mu\rightarrow \tilde{\mu}/a(t)$ \cite{gdea}.
Ressumation of the Schwinger DeWitt expansion in the large
curvature limit produce running couplings that depend on the
scalar curvature \cite{gdeR}.

In field theory, the running of the coupling constants is computed
in momentum space, and the relevant momentum scale is associated
to the characteristic energy in a scattering process. However, in
astrophysics and cosmology, one is eventually interested in the
running of the couplings in configuration space. As a trivial
property of Fourier transforms, a scale dependent coupling
constant in momentum state implies in general a non local
dependence in configuration space. Therefore, if the running of
couplings is produced by quantum effects (they could also be
produced by extra dimensions, see below), in principle one should
be able to compute a non local effective action and obtain real
and causal effective equations for the gravitational field.

This program can be explicitly worked out for weak gravitational
fields, in the case where quantum fields are free and
massless. Indeed, after integrating the quantum degrees of
freedom, the effective action \cite{barvilko} is equivalent to the
classical action with the replacement of the coupling constants by
the non local kernels \cite{einslang}
\begin{equation}
\alpha_i(\bar{\mu})\rightarrow \alpha_i(-\Box)=\alpha_{i\,
0}(\mu)+c_i\ln\left[-\frac{\Box}{\mu^2}\right] \label{rc1}
\end{equation}
where $\mu$ is a reference energy scale and the values of the
constants $c_i$ are fixed by the RGE. Therefore,  in this
particular example the running of the gravitational constants is
correctly and completely taken into account by replacing
$\bar\mu^2 =-\Box$ at the level of the effective action (this
approach has also been considered in  Refs. \cite{elizod,espriu}).
The effective field equations for the gravitational field give
rise to a quantum corrected Newtonian potential, that in turn can
be interpreted as the classical one with a running gravitational
constant.

As already stressed, if one assumes that the modifications to
General Relativity come from quantum effects, the running behavior
of the coupling constants should be contained in the effective
action \cite{qftrc}, along with other quantum corrections. The
``Wilsonian" approach, in any of the versions described above, is
a way of partially taking into account the effects of quantum
fields.
The running of the coupling constants is not necessarily
logarithmic. For example, renormalization group analysis based on the effective average action of
quantum Einstein gravity \cite{nonpert} and non perturbative studies in the lattice theory of quantum gravity \cite{latice},
suggests different behaviors for the scale dependence of the
gravitational constants, as for instance a power law scaling of
Newton's constant. One could implement this running in either
cosmological or astrophysical scenarios following the ideas
described above, by an adequate replacement of the momentum by
$r^{-1}$ or $t^{-1}$. A covariant approach that encompasses both
scenarios, based on the use of the kernel $G(-\Box)$ has also been
suggested \cite{hw1}. This is difficult to implement at the level
of the Einstein-Hilbert action, because formally any (analytic)
function of the d'Alembertian acting on the Ricci scalar is a
total derivative and does not contribute to the field equations.
When implemented at the level of  the equations of motion, the
modified Einstein equations take the form
\begin{equation}
R_{\mu\nu}-\frac{1}{2}R g_{\mu\nu}= 8\pi G(-\Box )T_{\mu\nu}\,\, .
\label{EEGvar}
\end{equation}

 For example, in the context of lattice theory of quantum gravity, it has been suggested that the strength of gravitational
 interactions might slowly increase with distance as a consequence of vacuum polarization effects. Provided that a non trivial ultraviolet
 fixed point exists, to leading order in the vicinity of such fixed point the scale dependence of the Newton's constant is characterized by a
 critical exponent $\nu=(2\alpha)^{-1}$, and the size of the corrections is set by a non perturbative scale $L$.  In this scenario,
 the manifestly covariant non local gravitational coupling  can be written as \cite{hw1}
\begin{equation}\label{runnLatt}
G(-\Box)=G_{N}\left(1+a_0 L^{-2\alpha}({-\Box+L^{-2}})^{-\alpha}\right),
\end{equation}
where $G_N$ is the usual Newton's constant, $a_0$ is a positive
number whose typical value was estimated to be of order of
$10^{-2}$, and estimates for the value of $\nu^{-1}=2\alpha$ vary
from $\nu^{-1}\approx 3.0$ to $\nu^{-1}\approx 1.7$ \cite{latice}.
The length scale $L$ separates the ultraviolet regime where non
perturbative corrections can be neglected from long distance
regime where such corrections become significant. This scale might
be related to the vacuum expectation value of the curvature of the
spacetime, and therefore, in cosmological situations, it is
generally chosen to have a typical value of order of the Hubble
ratio $H^{-1}$ \cite{hw1}.

Similar proposals have been considered to introduce infrared
corrections to General Relativity, which may help to understand
the present acceleration of the universe. These are in turn
inspired by extra dimensional brane world models (see below) \cite{dgp}.

Effective equations like Eq.(\ref{EEGvar}) are in general
inconsistent, i.e. the Bianchi identity is not satisfied. There is
a way out, which consists in considering the effective field
equations in the weak field limit. In this case, it is possible to
write a non local effective action whose variation gives the
effective field equations to linear order in the curvature
\cite{barvinsky}. This procedure is consistent, although
restricted to weak fields. A more drastic approach is to assume
that Eq.(\ref{EEGvar}) is valid to all orders, and that the
allowed energy-momentum tensors are only those that satisfy
$\left( G(-\Box)T^{\mu\nu}\right )_{;\nu}=0$.

Another problem of Eq.(\ref{EEGvar}) is that, unless the non local
corrections contained in $G(-\Box)$ are treated perturbatively,
the theory will contain ghosts. Indeed, as shown in
Ref.\cite{Dvali06}, at the linearized level the ghost-free
theories must be of the form
\begin{equation}
G_{\mu\nu}^{(1)} - m^2(-\Box )(h_{\mu\nu}-\eta_{\mu\nu} h)= 8\pi
G_N T_{\mu\nu} \label{fp}
\end{equation}
where $G_{\mu\nu}^{(1)}$ is the linearized Einsten's tensor,
$g_{\mu\nu}=\eta_{\mu\nu}+h_{\mu\nu}$ and
$h=\eta^{\mu\nu}h_{\mu\nu}$. Eq.(\ref{fp}) generalizes the
Pauli-Fierz theory of massive gravitons \cite{PF} to the case of a
non local mass $m^2(-\Box )$. Note that the linearization of
Eq.(\ref{EEGvar}) is not of the Pauli-Fierz form.

In brane world scenarios, the non local kernels previously considered in the literature
involve either arbitrary powers or the logarithm of the
d'Alembertian. For example, in order to modify General Relativity
at distances larger that a certain scale $L$, it has been assumed
that the leading behavior of the kernels is
\begin{equation}
\frac{G_N}{G(-\Box)}\sim \frac{m^2(-\Box)}{\Box}\sim
L^{-2\alpha}(-\Box )^{-\alpha}
\end{equation}
where $\alpha$ is a positive number \cite{bound} ($\alpha=1/2$ for the five dimensional DGP model \cite{dgp}) and $L$ is a length scale
which for cosmological situations is naturally assumed to be of order of  $H^{-1}$ . Note that in contrast to the running suggested by non
perturbative analysis of quantum effects (see Eq. \ref{runnLatt}), the Newton's constant here decrease with distance.  As an illustration,
a ``phenomenological" running coupling of the form \be
\frac{1}{G(-\Box)}= \frac{1}{G_N}(1+L^{-2\alpha}(-\Box + b^2
)^{-\alpha}) \label{defG} \ee with $bL\ll 1$, interpolates between
the usual Newton's constant at laboratory distance scales and $G(0)\ll G_N$
at scales much larger than $L$.

In order to analyze the effective field equations it is necessary
to define properly the action of an arbitrary power of the
d'Alembertian on scalar and tensor functions. This is one of the
main goals of the present paper. We will introduce a formal
definition based on an integral representation, as it has been
previously done with the logarithm of the d'Alembertian
\cite{barvilko}, and use this representation to obtain the action
of the non local operator on arbitrary functions. We will consider
both astrophysical and cosmological situations. This problem has
been previously considered by Hamber and Williams, who computed
$(-\Box)^{-\alpha}$ by analytic continuation of the action of
$(-\Box)^n$, for an integer number $n\geq 1$, to the case
$n=-\alpha\in \mathbb{R}$, in Robertson Walker \cite{hw1} and
Schwarzschild metrics \cite{hw2}. However, this approach has some
drawbacks. On the one hand, it is only possible to get an answer
for the action of $(-\Box)^{-\alpha}$ on some particular functions
that depend either on $t$ or $r$. On the other hand, the non local
and causal properties of the kernel are not apparent from its
definition.

The paper is organized as follows. In the next Section we
introduce our formal definition for $(-\Box)^{-\alpha}$. In
Section \ref{sec:flat} we calculate it explicitly in flat
spacetime,  and reobtain previous results based on a different
definition. We also show that the distribution $\ln[-\Box] $ can
be obtained as a particular limiting case. In Section
\ref{sec:FRW} we compute explicitly the action of
$(-\Box)^{-\alpha}$ on a scalar function independent of the
spatial coordinates in Friedman Robertson Walker (FRW) spacetimes.
We consider the de Sitter spacetime, and other FRW metrics with
initial singularities. We also study the action of the kernel
$(-\Box +b^2)^{-\alpha}$ on a constant function, and check whether
it gives $b^{-2\alpha}$ or not, a result that has been used to
argue that Eqs. like (\ref{EEGvar}) can explain the smallness of
the cosmological constant. In Section \ref{sec:log} we use the
results obtained for the kernel $(-\Box)^{-\alpha}$ to derive
similar ones for $\ln[-\Box]$. As a concrete application, in
Section \ref{sec:waves} we discuss the effect of a running
Newton's constant on the generation of gravitational waves. We
summarize our results in Section \ref{sec:conc}.

Throughout the paper we set $\hbar =c=1$ and adopt the sign
convention denoted (+++) by Misner, Thorne, and Wheeler
\cite{MTW}.

\section{Formal definition of $(-\Box)^{-\alpha}$}\label{sec:def}

We will define the kernel $(-\Box)^{-\alpha}$ using an integral
representation in terms of the two point function for massive
fields. The definition is based on the formal identity
\begin{equation}
(-\Box)^{-\alpha}=\frac{2 \sin(\alpha\pi)}{\pi} \int_0^{+\infty}
dm \; m^{1-2\alpha} \frac {1}{-\Box +m^2} \label{integralrep}
\end{equation}
valid for $0<\alpha <1$. A similar representation has been
previously considered \cite{barvilko} to define $\ln [-
\frac{\Box}{\mu^2}]$, i.e.
\begin{equation}
\ln \left[-\frac{\Box}{\mu^2}\right]=\int_0^{+\infty} dm^2 \left[-\frac
{1}{-\Box +m^2}+\frac {1}{\mu^2 +m^2}\right] \label{formaldeflog}
\end{equation}
It is worth noting that both kernels are related: knowing the
kernel $(-\frac{\Box}{\mu^2})^{-\alpha}$, it is possible to obtain
$\ln[-\frac{\Box}{\mu^2}]$ by considering the identity
$x^{-\alpha}=1-\alpha\ln x+O(\alpha^2)$ \cite{espriu}.

The tensorial character of the kernels is defined by the one of
the two point function of the massive field. When
$(-\Box)^{-\alpha}$ acts on a scalar (tensorial) function,
$({-\Box +m^2})^{-1}$ represents the propagator of a scalar
(tensorial) massive field. In what follows we will mainly consider
the action on scalar functions.

The definitions given above are unambiguous in Euclidean space,
where $-\Box +m^2$ is a unique positive definite operator. In
Lorentzian spacetimes, it is necessary to impose an additional
prescription to select one of the propagators $(-\Box +m^2)^{-1}$
(retarded, advanced, Feynman, etc). As we will be imposing the
modified field equations ``phenomenologically", i.e. without
considering the full underlying field theory, it is not possible
to give such a prescription. An educated guess is to impose the
field equation to be causal, and this can be implemented by
choosing the retarded propagator in the spectral representations.
This is the correct prescription when the effective equations come
from quantum corrections, and in particular for weak gravitational
fields. In this case, this choice corresponds to the vacuum
quantum state of the matter fields \cite{barvilko}. For the rest
of the paper we will follow this particular prescription, although
it is clear that it is not the only possibility. The scale
dependence of the Newton constant could be a consequence of the
presence of extra dimensions, and in this case it is not even
clear that the effective equations on the brane should be causal \cite{barvinsky}.
Moreover, even if the corrections come from matter quantum fields,
one could consider more general initial states. These issues
deserve further investigation.

Before addressing specific calculations, we introduce the
following notation for the kernels:  $(-\Box+b^2)^{-\alpha}\equiv
G_b^{\alpha}(x,x')$ and $\ln[-\frac{\Box}{\mu^2}]\equiv L(x,x')$.
Moreover, $G(M^2,x,x')$ will stand for the retarded propagator of
a scalar field of mass $M$. The action of the kernels on a scalar
function $f$ is explicitly given by \be
\ln\left[-\frac{\Box}{\mu^2}\right]f\equiv\int
d^4x'\sqrt{|g(x')|}L(x,x')f(x')\,\, ,\ee and \be\label{def}
(-\Box+b^2)^{-\alpha} f\equiv \int d^{4}x' \sqrt{|g(x')|}
G^{\alpha}_b(x,x') f(x'), \ee where \be\label{Galpha}
G^{\alpha}_b(x,x')=\frac{2 \sin(\alpha\pi)}{\pi} \int_0^{+\infty}
dm \; m^{1-2\alpha} G(m^2+b^2,x,x'). \ee

\section{$(-\Box)^{-\alpha} $ in Minkowski spacetime} \label{sec:flat}

Let us begin by computing the application of the operator defined
in  Eq.(\ref{def}) in Minkowski spacetime to an arbitrary
function. For this purpose, we  start by considering the retarded
propagator for a massive field, which is twice the real and causal
part of the Feynman propagator, \ba G(m^2+b^2,x,x')&=&2
\theta(t-t') \Re[G_F(x,x')]=2 \theta(t-t') \Re\left[\int
\frac{d^4k}{(2\pi)^4} \frac{\exp\{i k\cdot
x\}}{k^2-i\epsilon+m^2+b^2}\right]\\\nonumber &=& 2
\theta(t-t')\int_0^{+\infty} \frac{dk}{(2\pi)^2} k
\exp\{-\epsilon(t-t')\}\frac{\sin(k|\vec{x}-\vec{x'}|)}{|\vec{x}-\vec{x'}|}
\frac{\sin(\sqrt{k^2+m^2+b^2}(t-t'))}{\sqrt{k^2+m^2+b^2}}, \ea
where the positive number $\epsilon\to 0$ at the end of the
calculations.

The generalized Green's function $G^{\alpha}_b(x,x')$ can be then
obtained by inserting this result into Eq.(\ref{Galpha}) and
performing the remaining integration. In the case in which $b=0$
we get \cite{GR}
\be\label{Minkparatodo} G^{\alpha}_0(x,x') =
\frac{\Gamma(1/2)\Gamma(\alpha+1/2)}{\pi^2\Gamma(\alpha-1)\Gamma(2\alpha)}
\theta(t-t'-|\vec{x}-\vec{x'}|)[(t-t')^2-|\vec{x}-\vec{x'}|^2]^{\alpha-2}
\exp\{-\epsilon(t-t')\}, \ee which is the same as the
corresponding retarded  Green's function obtained by Bollini and
Giambiagi in Ref.\cite{BollGiambi} using a different method. Note that this generalized
Green's function does not satisfy the Huygens' principle, i.e. its
support is not restricted to the surface of the past light cone.

With this result we can apply the $(-\Box)^{-\alpha}$ operator to
any test function that depends on the spacetime coordinates. In
the case of time independent functions, the time integral of the
retarded propagator coincides with the Green's function of the
Laplacian, \be\int_{-\infty}^{+\infty} dt' G(m^2+b^2,x,x')=\int
\frac{d^3k}{(2\pi)^3}\frac{\exp\{i\vec{k}\cdot(\vec{x}-\vec{x'})\}}{k^2+m^2+b^2}.\ee
Therefore, in such cases we obtain \ba (-\Box + b^2)^{-\alpha}
f&=&\int d^3x' \int
\frac{d^3k}{(2\pi)^3}\frac{\exp\{i\vec{k}\cdot(\vec{x}-\vec{x'})\}}{(k^2+b^2)^{\alpha}}
f(\vec{x'})\\\nonumber &=&\int
\frac{d^3k}{(2\pi)^3}\frac{\exp\{i\vec{k}\cdot\vec{x}\}}{(k^2+b^2)^{\alpha}}
f(\vec{k}), \ea where $f(\vec{k})=\int d^3x
\exp\{-i\vec{k}\cdot\vec{x}\} f(\vec{x})$.

As one can immediately note from the equation above, the
application of $(-\Box+b^2)^{-\alpha}$ to a constant function
$f=1$ gives $b^{-2\alpha}$. This result is the one that would be
expected by expanding $(-\Box+b^2)^{-\alpha}$ in powers of the
derivative operator $\Box$.

On the other hand, if the test function only depends on the time
coordinate we are allowed to perform the integration on the space
coordinates, with the result  \be \label{gMinkconb}
g^{\alpha}_b(t,t')\equiv\int d^3x' G^{\alpha}_b(x,x')=
2^{1/2-\alpha} b^{1/2-\alpha}\frac{\Gamma(1/2)}{\Gamma(\alpha)}
J_{\alpha-1/2}(b(t-t'))\theta(t-t')(t-t')^{\alpha-1/2}, \ee where
$J$ is the Bessel function of the first kind \cite{Abram}. For
$b=0$ it reduces to \be\label{gMink}
g^{\alpha}_0(t,t')=\frac{\exp\{-\epsilon(t-t')\}}{{\Gamma(2\alpha)}}\theta(t-t')(t-t')^{2\alpha-1}.\ee

As particular examples, let us consider the test functions
$f_1(t)=(-t)^{\beta}$ $(t<0,\beta<-2\alpha)$ and $f_2(t)=\exp\{\nu
t\}$ $(\nu>0)$, for which we obtain \ba \label{minctalabeta}
(-\Box)^{-\alpha}f_1&=&\int_{-\infty}^{+\infty} dt'
g^{\alpha}_0(t,t')
f_1(t')=\frac{\Gamma(-2\alpha-\beta)}{\Gamma(-\beta)}(-t)^{2\alpha+\beta},\\
(-\Box)^{-\alpha}f_2&=&\int_{-\infty}^{+\infty} dt'
g^{\alpha}_0(t,t') f_2(t')=\nu^{-2\alpha}\exp\{\nu t\}.
 \ea
Note that as the definition we are considering is not applicable
to all integer values of $\beta$, one cannot obtain the second
result from the first one by a Taylor expansion of the exponential
function. Note also that both results can be obtained from
analytic continuation, by considering the application of the
derivative operator $(-\Box)^{n}$ with $n\in \mathbb{N}$ that
yields a real result after replacing n by $-\alpha$
\cite{Zavadaetc}.

The results of this Section are useful to analyze field equations
like Eq.(\ref{EEGvar}) in the weak field limit. We will use them
in Section \ref{sec:waves}, to study the generation of
gravitational waves in the linearized limit.

\section{$(-\Box)^{-\alpha} $ in FRW spacetimes} \label{sec:FRW}

In this Section we consider three qualitatively different
spatially flat FRW spacetimes. We use coordinates such that \be
ds^2 = a^2(\eta ) \left[-d\eta^2 + \sum_{i=1}^3 dx_i^2 \right]=
-dt^2+a^2(t)\sum_{i=1}^3 dx_i^2 ,\ee where $a$ is the scale factor,
$\eta$ is the conformal time ($d\eta = dt/a(t)$) and  $t$ the
cosmic time.

\subsection{De Sitter spacetime}

We set the scalar factor to be $a(\eta)=-(H\eta)^{-1}$, where the
constant $H=\dot{a}/a$ (with a dot denoting differentiation with
respect to $t$). The retarded propagator  for massive scalar field
in this spacetime is given by \cite{birrell} \be
G(m^2+b^2,x,x')=\theta(\eta-\eta')\Re\left\{i \frac{H^2}{8\pi}
\frac{(1/4-\nu^2)}{\cos(\pi\nu)}F\left[\frac{3}{2}+\nu,\frac{3}{2}-\nu,2;
1+\frac{(\eta-\eta'-i\epsilon)^2-|\vec{x}-\vec{x'}|^2}{4\eta\eta'}\right]\right\},\ee
where $F$ is the Gauss's hypergeometric function \cite{Abram} and
$\nu^2=9/4-(m^2+b^2)/H^2$.

As in this case the integration on the mass parameter $m$ is very
difficult, we restrict ourselves to test functions that depend
only on the time coordinate. Then, before performing the
integration on the mass we integrate on the space coordinates,
which results in \be \label{deSitterTimeProp}
g(m^2+b^2,\eta,\eta')\equiv\int d^3x' \sqrt{|g(\eta')|}
G(m^2+b^2,x,x')=\frac{\theta(\eta-\eta')}{H^2\nu}
\frac{(-\eta)^{3/2}}{{(-\eta')}^{5/2}}\sinh\left(\nu\ln\left(\frac{\eta'}{\eta}\right)\right).\ee
This equation can be integrated on the mass parameter to find \be
g_b^{\alpha}(\eta,\eta')=\frac{2 \sin(\alpha\pi)}{\pi}
\int_0^{+\infty} dm \; m^{1-2\alpha} g(m^2+b^2,\eta,\eta').\ee The
result depends on whether $\nu$ is a real or purely imaginary
number. We get \ba\label{deSitterRes}
g_b^{\alpha}(\eta,\eta')&=&\theta(\eta-\eta')
H^{-2\alpha}\frac{\Gamma(1/2)}{\Gamma(\alpha)}\frac{(-\eta)^{3/2}}{(-\eta')^{5/2}}
\left(\frac{1}{2}\ln\left(\frac{\eta'}{\eta}\right)\right)^{\alpha-1/2}\\\nonumber
&\times&\left\{%
\begin{array}{ll}
\left(\frac{9}{4}-\frac{b^2}{H^2}\right)^{1/4-\alpha/2}\;
I_{\alpha-\frac{1}{2}}\left(\sqrt{\frac{9}{4}-\frac{b^2}{H^2}}\ln\left(\frac{\eta'}{\eta}\right)\right),
&
\hbox{$b^2\leq 9/4 H^2\;$;} \\
\left(\frac{b^2}{H^2}-\frac{9}{4}\right)^{1/4-\alpha/2}\;
J_{\alpha-\frac{1}{2}}\left(\sqrt{\frac{b^2}{H^2}-\frac{9}{4}}\ln\left(\frac{\eta'}{\eta}\right)\right),
&
\hbox{$b^2\geq 9/4 H^2\;$;} \\
\end{array}%
\right.\ea where $I$ is the modified Bessel function of the first
kind \cite{Abram}.

As in Minkowski spacetime, applying the operator
$(-\Box+b^2)^{-\alpha}$ to a constant $f=1$ we  obtain the
``expected" result $b^{-2\alpha}$. As we will see in the next
sections, this property is not true in spacetimes with initial
singularities. This is important, for example, in the context of
models where in order to explain the smallness of the
experimentally inferred dark energy density, the conventional laws
of Einstein's gravity are modified at very large scales by
introducing non local operators. In such models, gravity is
modified so that a large vacuum energy density (identified with
the dark energy density) does not give rise to large observable
curvature. Phenomenologically, it can be seen as a result of the
possibility of having a modified Einsten's equation like
Eq.(\ref{EEGvar}) with the non local effective gravitational
coupling such that when it is applied to a smooth and persistent
source yields a small contribution due to $G(\Box)1=G(0)<<G_N$.
For example, in the presence of a vacuum energy density,
$T_{\mu\nu}=-\rho_v g_{\mu\nu}$, the trace of Eq.(\ref{EEGvar})
becomes
\begin{equation}
R=32\pi G(-\Box)\rho_v, \end{equation} so the effective vacuum
energy is expected to be $G(0)\rho_v\ll G_N\rho_v$. As a
phenomenological approach, the idea of addressing the cosmological
problem with non local modification of gravity in the infrared is
discussed for example in Ref.\cite{DGPAcaussal}, where it is
argued that acausality is a fundamental feature for this program
to work.

The application of $(-\Box)^{-\alpha}$ to $f(\eta)=(-H\eta)^k$
($k<0$) yields \be\label{testetaalak} \int d\eta'\;
g^{\alpha}_0(\eta,\eta')f(\eta')=\frac{(-H\eta)^{k}}{
[k(k-3)]^{\alpha}H^{2\alpha}},\ee which can also be seen as an
analytic continuation of the result of the application of the
covariant derivative operator $(-\Box)^{n}$, with $n$ an integer
number, from $n\to -\alpha$ (we remind that we are assuming
$0<\alpha <1$).

It is useful to note that if we are only interested in applying
the kernel $(-\Box+b^2)^{-\alpha}$ to test functions that are
independent on the space coordinates we only need to know the zero
mode of the scalar field. That is, we can start by working out the
solution of the Klein-Gordon equation for a spatially homogeneous
scalar field $\phi(\eta)=(2\pi)^{-3/2} \chi(\eta)/a(\eta)$ (with
mass $M=(m^2+b)^{1/2}$ and minimally coupled to gravity),
\be\label{KGeqconb}
\chi''(\eta)+a^2(\eta)\left[m^2+b^2-\frac{R}{6}\right]\chi(\eta)=0,
\ee where a prime means a derivative with respect to $\eta$ and $R=6
a''(\eta)/a^3(\eta)$ is the Ricci scalar. After imposing the
Wronskian normalization condition,
\be\label{wronskian}\chi(\eta)\partial_\eta\chi^{*}(\eta)-\chi^{*}(\eta)\partial_\eta\chi(\eta)=i,\ee
the retarded propagator integrated in the space coordinates is
determined by \be\label{retpropzeromode}
g(m^2+b^2,\eta,\eta')\equiv\int d^3x \sqrt{|g(\eta')|}
G(m^2+b^2,x,x')=2\theta(\eta-\eta')\sqrt{|g(\eta')|}\Re\left\{i
(2\pi)^3\phi(\eta)\phi^{*}(\eta')\right\}.\ee We can reobtain the
result presented in Eq.(\ref{deSitterTimeProp}) by starting with
the following solution to Eq.(\ref{KGeqconb}):
\be\label{deSitterZeroMode}
\phi(\eta)=\frac{\chi(\eta)}{a(\eta)(2\pi)^{3/2}}=\frac{(-\eta)^{1/2}}{a(\eta)
(2\pi)^{3/2}}\left\{%
\begin{array}{ll}
\frac{(-\eta)^{\nu}-i(-\eta)^{-\nu}}{2\sqrt{\nu}}, & \hbox{$\nu\in\mathbb{R}$;} \\
\frac{(-\eta)^{\nu}}{\sqrt{2 |\nu|}}, & \hbox{$i\nu\in \mathbb{R}$.} \\
\end{array}%
\right.\ee As one can easily check, this solution satisfies the
Wronskian condition (\ref{wronskian}). Then, in order to find the
propagator $g(m^2+b^2,\eta,\eta')$ given in
Eq.(\ref{deSitterTimeProp}) we only need to insert this solution
into Eq.(\ref{retpropzeromode}). It is worth noting that the
result does not depend on the initial condition for the mode.

For the remaining FRW spacetimes considered below, we restrict
ourselves to test functions that depend only on time, and therefore we
will follow this last procedure.

\subsection{FRW spacetime with $a(t)\varpropto t^{1/2}$}

We now consider a radiation dominated FRW spacetime with scale
factor  $a(t)= a_0 t^{1/2}$. This spacetime has an initial
singularity and a finite particle horizon.

A solution of Eq.(\ref{KGeqconb}) that satisfies the Wronskian
normalization condition is given by \be
\phi(\eta)=\frac{\sqrt{\eta}}{8 \pi a(\eta) } H_{1/4}^{(2)}(M t),
\ee where $t=a_0^2\eta^2/4$, $M=\sqrt{m^2+b^2}$, and $H^{(2)}$ is
the Hankel function of the second kind \cite{Abram}. Substituting
this solution into Eq.(\ref{retpropzeromode}) we obtain \be
g(m^2+b^2,\eta,\eta')= \theta(\eta-\eta')\frac{\pi
a_0^4}{2^{9/2}}\frac{{\eta'}^{7/2}}{\eta^{1/2}} \left\{J_{1/4}(M
t)J_{-1/4}(M t')- J_{-1/4}(M t)J_{1/4}(M t')\right\}. \ee
 For $b=0$ the mass integration yields
\be\label{gRadiac}
g^{\alpha}_{0}(\eta,\eta')=\theta(\eta-\eta')\frac{t^{2\alpha}2^{2-2\alpha}}{\Gamma(2\alpha)}
 \frac{{\eta'}^3}{\eta^4}
\left(1-\frac{{\eta'}^4}{\eta^4}\right)^{2\alpha-1}
F\left[\alpha-\frac{1}{4},\alpha,2\alpha;
1-\frac{{\eta'}^4}{\eta^4}\right].\ee

Contrarily to the previous cases, after applying the
$(-\Box)^{-\alpha}$ operator to a constant $f=1$ we obtain a
finite result, \be \int_{0}^{+\infty} d\eta'\;
g^{\alpha}_{0}(\eta,\eta')=\frac{\Gamma(5/4)t^{2\alpha}
4^{-\alpha}}{\Gamma(\alpha+1)\Gamma(5/4+\alpha)}. \ee This finite
result is a consequence of the fact that this spacetime has an
initial singularity since, as it was pointed out in
Ref.\cite{DGPAcaussal}, starting at $t=0$ ($\eta=0$ in this case)
it would take causal physics too much time to recognize that the
constant was truly a constant. This interpretation may be further
supported by computing the application of the
$(-\Box+b^2)^{-\alpha}$ to $f=1$, \be \int_{0}^{+\infty} d\eta'\;
g^{\alpha}_{b}(\eta,\eta')=\sin(\pi\alpha)t^{2\alpha}2^{-2\alpha+1/2}
\frac{\Gamma(-\alpha)}{\Gamma(-1/4)\Gamma(5/4+\alpha)}
{}_1F_2\left[\alpha,1+\alpha,\frac{5}{4}+\alpha;-\frac{b^2t^2}{4}\right],
\ee where ${}_1F_2$ is one of the generalized hypergeometric
function ${}_pF_q$ \cite{wolfram}. Then, taking the limit $t\to
+\infty$ we arrive at the expected result \be \lim_{t\to
+\infty}\int_{0}^{+\infty}
 d\eta'\; g^{\alpha}_{b}(\eta,\eta')=b^{-2\alpha}. \ee

On the other hand, despite the initial singularity, applying
$(-\Box)^{-\alpha}$ to
$f(\eta)=(a_{0}^2\eta^2/4)^{\beta}=t^{\beta}$ ($\beta>-2$) we get
\be \label{talabetarad} \int_{0}^{+\infty} d\eta'\;
g^{\alpha}_{0}(\eta,\eta')
f(\eta')=4^{-\alpha}t^{2\alpha+\beta}\frac{\Gamma(\beta/2+1)\Gamma(\beta/2+5/4)
}{\Gamma(\beta/2+\alpha+1)\Gamma(\beta/2+5/4+\alpha)}\,\, . \ee
This result can also be obtained through analytic continuation of
the application of covariant derivative operator $(-\Box)^n$ for
positive integers $n$ \cite{hw1}.

\subsection{FRW spacetime with $a(t)\varpropto t^2$}

We will now consider a FRW spacetime with an initial singularity
but, in contrast to the previous example, without a finite
particle horizon. The choice $a(t)= a_0 t^{2}$ will allow us to
find a solution of Eq.(\ref{KGeqconb}) in terms of elementary
functions. Indeed, a solution that satisfies the Wronskian
condition in Eq.(\ref{wronskian}) is given by \be
\phi(\eta)=\frac{\exp\{-i M t\}}{a(\eta)(2\pi)^{3/2}(M
t)^{3}}\sqrt{\frac{M}{2 a_0}}[M^2 t^2-3(1+i M t)], \ee with
$t=-(a_0 \eta)^{-1}$ and $M=\sqrt{m^2+b^2}$. Inserting this
solution into Eq.(\ref{retpropzeromode}) we find \ba
g(m^2+b^2,\eta,\eta')&=&\frac{\theta(\eta-\eta')}{a_0 M^5 t^3
{t'}^3} \frac{a^3(\eta')}{a(\eta)}\left\{\cos(M(t-t'))\left[3 M t
(M^2 {t'}^2-3)-3 M t'(M^2 t^2-3)\right]\right.\\\nonumber
&+&\left.\sin(M(t-t'))\left[9 M^2 t t'+(M^2t^2-3)(M^2
{t'}^2-3)\right]\right\},\ea and for $b=0$, after integrating on
the mass parameter, we obtain \be\label{gtcuad}
g^{\alpha}_0(\eta,\eta')=
\frac{a(\eta')\theta(\eta-\eta')(t-t')^{2\alpha-1}}{\Gamma(2\alpha)(1+2\alpha)(3+2\alpha)}
\left[3\frac{ t'}{t}+3\frac{{t'}^5}{t^5}+3(2\alpha-1)
\left(\frac{{t'}^2}{t^2}+\frac{{t'}^4}{
t^4}\right)+(3+4\alpha(\alpha-1))\frac{{t'}^3}{t^3}\right].\ee

The application to a constant $f=1$ yields the result \be
\int d\eta'\;
g^{\alpha}_0(\eta,\eta')=\frac{15
t^{2\alpha}}{\Gamma(2\alpha+2)(2\alpha+3)(2\alpha+5)}, \ee which
is also finite.

As in the previous case, applying $(-\Box+b^2)^{-\alpha}$ to a
constant we get
 \ba \int d\eta'\;
g^{\alpha}_b(\eta,\eta')&=&\frac{15}{\pi}t^{2\alpha}\sin(2\pi\alpha)
\left\{\Gamma(-3-2\alpha)
{}_1F_2\left[\alpha,2+\alpha,\frac{5}{2}+\alpha;-\frac{b^2
t^2}{4}\right]\right.\\\nonumber &+&\left. 3\Gamma(-4-2\alpha)
{}_1F_2\left[\alpha,\frac{5}{2}+\alpha,3+\alpha;-\frac{b^2
t^2}{4}\right]+3\Gamma(-5-2\alpha)
{}_1F_2\left[\alpha,3+\alpha,\frac{7}{2}+\alpha;-\frac{b^2
t^2}{4}\right]\right\}, \ea and taking the limit $t\to +\infty$ we
find \be \lim_{t\to+\infty}\int d\eta'\;
g^{\alpha}_b(\eta,\eta')=b^{-2\alpha}. \ee

Therefore, the fact that the application of the
$(-\Box)^{-\alpha}$ to a constant yields a finite result seems not
related to the  existence of a particle horizon, but to the
existence of a singularity at $t=0$.

Another interesting property we can point out is that considering
the advanced propagator, \be\label{Avpropzeromode}
g_{adv}(m^2+b^2,\eta,\eta')=-2\theta(\eta'-\eta)\sqrt{|g(\eta')|}\Re\left\{i
(2\pi)^3\phi(\eta)\phi^{*}(\eta')\right\},\ee instead of the
retarded one, we obtain \be (-\Box+b^2)_{adv}^{-\alpha}1 =
b^{-2\alpha}\,\, , \ee independently of the value of the initial
time $t$.

Finally, for the sake of completeness, we compute the application
of $(-\Box)^{-\alpha}$ to $f(\eta)=(-a_0 \eta)^{-\beta}=t^{\beta}$
($\beta>-2$), with the result
 \be\label{talaalphatcuad}\int d\eta'\; g^{\alpha}_b(\eta,\eta')f(\eta')=4^{-\alpha}
 t^{2\alpha+\beta}\frac{\Gamma(\beta/2+7/2)
\Gamma(1+\beta/2)}{\Gamma(1+\alpha+\beta/2)\Gamma(\alpha+7/2+\beta/2)},\ee
which can also be obtained by analytic continuation from an
integer number $n$ to $-\alpha$.

\section{The operator $\ln[-\Box/\mu^2]$} \label{sec:log}

In this Section we compute the application of the operator that
can represent the logarithm of the covariant d'Alembertian to
scalar test functions that are independent of the space
coordinates, for each of the four spacetimes considered above. As
described in Section \ref{sec:def}, we will do that by taking the
limit $\alpha\rightarrow 0$ in the kernel $(-\Box)^{-\alpha}$.

We begin with Minkowski spacetime. We start by applying the kernel
$(-\Box)^{-\alpha}$, given in Eq.(\ref{gMink}), to a generic test
function $f$ that is independent on the space coordinates, \be
\int_{-\infty}^{+\infty}dt'\; \mu^{2\alpha} g^{\alpha}_0 (t,t')
f(t')=\frac{1}{\Gamma(2\alpha)}\int_{-\infty}^t dt'\;
\mu^{2\alpha} \exp\{-\epsilon(t-t')\} (t-t')^{2\alpha-1} f(t'),\ee
where we have added a constant $\mu$ with mass dimensionality.

Before expanding in powers of $\alpha$, and in order to exclude explicitly the
local contribution, we perform an integration by
parts,   \be \frac{1}{\Gamma(2\alpha)}\int_{-\infty}^t dt'\;
\mu^{2\alpha} \frac{\exp\{-\epsilon(t-t')\}}{ (t-t')^{1-2\alpha}} f(t')=
f(t)
\left(\frac{\mu}{\epsilon}\right)^{2\alpha}-\frac{1}{\Gamma(2\alpha)}\int_{-\infty}^t
dt'\; \mu^{2\alpha}  \Gamma(2\alpha, \epsilon(t-t'))
\frac{d}{dt'}f(t'),\ee where $\Gamma(a,x)$ is the incomplete gamma
function \cite{Abram}, and we have assumed that
$f(t')\Gamma(2\alpha,\epsilon(t-t'))$ goes to zero as $t'\to
-\infty$.

After subtracting the local zeroth order contribution $f(t)$ and
expanding up to first order in $\alpha$ we find \be
\int_{-\infty}^{+\infty}dt'\; g^{\alpha}_0 (t,t') f(t')-f(t)\simeq
-\alpha\left\{\ln\left(\frac{\epsilon^2}{\mu^2}\right)+2
\int_{-\infty}^t dt' \Gamma(0,\epsilon(t-t'))
\frac{d}{dt'}f(t')\right\}.\ee

Finally, in the limit $\epsilon\to 0$ we obtain \be
\ln\left[-\frac{\Box}{\mu^2}\right]f=-2\gamma
f(t)-2\int_{-\infty}^t dt'\;\ln\left(\mu(t-t')\right)\frac{d}{d
t'} f(t'), \ee  where $\gamma$ is the Euler's constant and we have
assumed that $f(t')\to 0$ as $t'\to -\infty$. The same result can
also be obtained from the representation proposed for this kernel
in Ref.\cite{Jordan}.

With the use of this integral representation one can
straightforwardly compute the application of
$\ln[-\frac{\Box}{\mu^2}]$ to spatially homogeneous functions. For
example, for $f(t)=(-t)^{\beta}$ (with $t<0$ and $\beta<-2\alpha$)
it yields \be
\ln\left[-\frac{\Box}{\mu^2}\right]f=(-t)^{\beta}\{-\ln(t^2\mu^2)+2\psi(-\beta)\},
\ee where $\psi(x)$ is the Psi (Digamma) function \cite{Abram}.
Note that it can also be obtained by expanding the result in
Eq.(\ref{minctalabeta}) in powers of $\alpha$ and extracting the
first order contribution.

For the three remaining spacetimes, we can follow the same
procedure. Details of the calculations are relegated to the
Appendix \ref{Appendix}.

For the de Sitter spacetime, we arrive at \be\label{LNdeSitter}
\ln\left[-\frac{\Box}{\mu^2}\right]f
=\ln\left(\frac{H^2}{\mu^2}\right)f(\eta)-\int_{-\infty}^{\eta}
d\eta'\;\left\{\ln\left(\frac{e^{\gamma}}{3}\ln\left(\frac{\eta'}{\eta}\right)\right)
+E_i\left(-3\ln\left(\frac{\eta'}{\eta}\right)\right)\right\}\frac{d}{d
\eta'} f(\eta'), \ee where $E_i$ is the exponential-integral
function \cite{GR} and we have assumed that $f(\eta)\to 0$ as $\eta\to -\infty$. The
application of this operator to $f(\eta)=(-H\eta)^k$ ($k<0$) gives
the result \be\label{ResLNdeSitter}
\ln\left[-\frac{\Box}{\mu^2}\right]f =(-H\eta)^k
\ln\left(\frac{H^2}{\mu^2}k(k-3)\right). \ee

Likewise, for the radiation dominated FRW spacetime
($a(t)\varpropto t^{1/2}$) we obtain \ba \label{LNinRad}
\ln\left[-\frac{\Box}{\mu^2}\right]f&=&\left[4-2\gamma-\frac{\pi}{2}-\ln\left(\frac{
\mu^2}{2H^2}\right)\right]f(\eta)+\int_{0}^{\eta}
d\eta'\;\left\{\frac{4{\eta'}^5}{5\eta^5}
F\left[1,\frac{5}{4},\frac{9}{4};\frac{{\eta'}^4}{\eta^4}\right]\right.\\\nonumber
&-&\left.\ln\left(1-\frac{{\eta'}^4}{\eta^4}\right)\right\}\frac{d}{d
\eta'} f(\eta'), \ea with $f(\eta)\to 0$ as $\eta\to 0$. After
applying it to $f=t^{\beta}=(a_0\eta/2)^{2\beta}$ ($\beta>0$) we
get \be \label{LNtbetarad}
\ln\left[-\frac{\Box}{\mu^2}\right]f=t^{\beta}\left\{\ln\left(\frac{16
H^2}{\mu^2}\right)+\psi\left(\frac{\beta}{2}+\frac{5}{4}\right)+\psi\left(\frac{\beta}{2}+1\right)\right\}.\ee

Lastly, in the case of the FRW spacetime with $a(t)\varpropto
t^{2}$ the result is
 \be \label{LNintcuad}
\ln\left[-\frac{\Box}{\mu^2}\right]f=\left[\frac{46}{15}-2\gamma-\ln\left(
\frac{4\mu^2}{H^2}\right)\right]f(\eta)-\int_{0}^{t}
dt'\;\left\{\frac{2{t'}^5}{5 t^5}+\frac{2 t'}{ t}+\frac{2{t'}^3}{3
t^3} +2\ln\left(1-\frac{{t'}}{t}\right)\right\}\frac{d}{d t'}
f(t'), \ee  with $f(t)\to 0$ as $t\to 0$. For
$f=t^{\beta}=(-a_0\eta)^{-\beta}$ ($\beta>0$) it yields \be
\label{lntbetatcuad}
\ln\left[-\frac{\Box}{\mu^2}\right]f=t^{\beta}\left\{\ln\left(\frac{H^2}{\mu^2}\right)
+\psi\left(\frac{\beta}{2}+\frac{7}{2}\right)+\psi\left(\frac{\beta}{2}+1\right)\right\}.\ee

As we have pointed out in the Introduction, there are different
proposals in the literature to take into account the running of
coupling constants, which correspond to different replacements of
the energy scale $\bar{\mu}$ by $1/t$, $H$, $\tilde\mu/a(t)$, etc.
In cosmological situations, it is common to assume a power law
behavior for both the scale factor and the matter content. In such
situations, the relevant results are the ones given in
Eqs.(\ref{talabetarad}), (\ref{talaalphatcuad}),
(\ref{LNtbetarad}), and (\ref{lntbetatcuad}), which were obtained
by applying the corresponding non local kernel to test functions
that have a power law dependence on the time coordinate (either
$\eta$ or $t$).  These results can be written in the form
\ba\label{genera}
 (-\Box)^{-\alpha}f&\varpropto& H^{-2\alpha}f,\\
 \ln\left[-\frac{\Box}{\mu^2}\right]f
&=&\ln\left(\frac{H^2}{\mu^2}\right)f + const\times f\,\,. \ea
Moreover, in de Sitter spacetime, if we consider test functions
that have a power law dependence on the conformal time $\eta$, we
can also write the results in the same form (see
Eqs.(\ref{testetaalak}) and (\ref{ResLNdeSitter})). This would
suggest that in FRW spacetimes the action of the non local kernel
is equivalent to the replacement $\bar{\mu}\to H$. However, this
is not strictly correct. Indeed, the results depend on the test
function. On the one hand, the constant of proportionality in Eq.(\ref{genera}) depends on $\alpha$ and $\beta$. On the other hand,
in de Sitter spacetime, if we
apply the non local operator to a function $f=(-t)^{\beta}$ ($\beta<0$ and $t<0$)  we
obtain \be\label{DeSittertbeta}
\ln\left[-\frac{\Box}{\mu^2}\right]f
=(-t)^{\beta}\left\{\ln\left(\frac{H^2}{\mu^2}\right)+F(Ht,\beta)\right\},
\ee where  $F(Ht,\beta)$ is a complicated function whose
functional dependence on time is not logarithmic. For example, in
the particular case of $\beta=-2$ it reduces to \be
F(Ht,-2)=2-\gamma-\ln(-Ht/3)-\exp\{-3Ht\}(1+3Ht)E_i(3Ht). \ee This
non logarithmic behavior may have been expected from the fact
that, in contrast to the other FRW examples, for de Sitter
spacetime one has $t$ and $H$ as two different and relevant
scales.

\section{Generation of gravitational waves}\label{sec:waves}

In this Section we analyze the modified Einstein's equations
(\ref{EEGvar}) with a running coupling given by Eq. (\ref{defG}),
for the particular case $b=0$. We assume weak gravitational
fields. Using an expansion in powers of the curvature, it can be
shown that the modified equations follow from the non local action
\cite{barvinsky} \be S=-\int d^4x \sqrt{|g|}\left\{\frac{1}{16\pi
G_N}\left(R^{\mu\nu}-\frac{1}{2}g^{\mu\nu}R\right)\frac{1+F[-\Box]}{\Box}R_{\mu\nu}+
\mathcal{O}(R_{\mu\nu}^3)\right\}+S_{matter},\ee where
$S_{matter}$ corresponds to the source of the gravitational waves
($2 \delta S_{matter}=\sqrt{|g|} T_{\mu\nu} \delta g^{\mu\nu}$).
Here, the kernel $F[-\Box]=(-L^2\Box)^{-\alpha}$ represents a
small modification of Einstein's theory for very large scales.

As anticipated, we work in the weak field limit
$g_{\mu\nu}=\eta_{\mu\nu}+h_{\mu\nu}$, where $h_{\mu\nu}$
describes a linear perturbation to the Minkowski metric. In the
harmonic gauge, up to first order in $h_{\mu\nu}$, the modified
Einstein's equations become \be
(1+F[-\Box])\Box\bar{h}_{\mu\nu}=-16 \pi G_N T_{\mu\nu}, \ee where
$\bar{h}_{\mu\nu}=h_{\mu\nu}-\eta_{\mu\nu}h^{\gamma}_{\gamma}/2$.

We assume that the inclusion of the kernel $F[-\Box]$ yields a
perturbatively small correction to the standard result. Therefore,
we split
$\bar{h}_{\mu\nu}=\bar{h}_{\mu\nu}^{(0)}+\bar{h}_{\mu\nu}^{(1)}$,
where $\bar{h}_{\mu\nu}^{(0)}$ satisfy \be
\Box\bar{h}_{\mu\nu}^{(0)}=-16 \pi G_N T_{\mu\nu}, \ee and
$\bar{h}_{\mu\nu}^{(1)}$ can be perturbatively computed by solving
the following equation:\be
\Box\bar{h}_{\mu\nu}^{(1)}=-(-L^2\Box)^{-\alpha}\Box\bar{h}_{\mu\nu}^{(0)}=16
\pi G_N (-L^2\Box)^{-\alpha} T_{\mu\nu}\equiv -16 \pi G_N
\tilde{T}_{\mu\nu}.\ee

With the use of the retarded Green's function we can write
\be\bar{h}_{\mu\nu}^{(1)}(t,\vec{x})=4 G_N \int
\frac{d^3x'}{|\vec{x}-\vec{x'}|}
\tilde{T}_{\mu\nu}(t-|\vec{x}-\vec{x'}|,\vec{x'}), \ee where
\be\label{Ttilde}\tilde{T}_{\mu\nu}(t,\vec{x})=-\int d^4x'
G_0^{\alpha}(x,x') T_{\mu\nu}(x'). \ee Note that the value of
$\bar{h}_{\mu\nu}^{(1)}$ in the spacetime point $x$ is determined
only by the values that $\tilde{T}_{\mu\nu}$ takes on the past
light cone of $x$, but in order to compute $\tilde{T}_{\mu\nu}$
one needs to  know  all the values of the energy-momentum tensor
$T_{\mu\nu}$ inside that  past light cone. Therefore, the solution
for the metric perturbation $\bar{h}_{\mu\nu}$  does not obey the
Huygens' principle. This last issue was analyzed in detail in
Ref.\cite{BollGiambi} for different spacetime dimensions.

Finally, working with the Fourier transform and using the
generalized Green's function of Eq.(\ref{Minkparatodo}) one can
find \be \bar{h}_{\mu\nu}(\omega,\vec{k})=-16\pi
G_N\{1-L^{-2\alpha}[-(\omega+i\epsilon)^2+|\vec{k}|^2]^{-\alpha}\}
\frac{{T}_{\mu\nu}(\omega,\vec{k})}
{[-(\omega+i\epsilon)^2+|\vec{k}|^2]}.\ee

The non locality of the modified Einstein equations has other
interesting consequences. It gives rise to corrections to the Newtonian
potential which in turn imply that the metric outside a massive object may
depend on its inner structure, violating Birkhoff theorem
\cite{satz}. Therefore, we expect the corrections to the metric
outside a collapsing star to be time dependent, even in
spherically symmetric situations.

\section{Conclusions}\label{sec:conc}

We have presented a detailed derivation of the action of the non
local kernels $(-\Box)^{-\alpha}$ and $\ln[-\Box]$ on scalar
functions. Our starting point was the representation of both
kernels in terms of integrals of the retarded propagator of a
massive scalar field.

In Minkowski spacetime, we derived an explicit expression for the
kernel  $(-\Box)^{-\alpha}$. This is relevant for the analysis of
the modified Einstein Eq.(\ref{EEGvar}) in the weak field limit.
As an application, we obtained formal expressions for the
generation of gravitational waves, and showed explicitly that
Huygens' principle is not longer valid when one considers a scale
dependence of Newton constant.

In Robertson Walker spacetimes, we considered the action of
$(-\Box)^{-\alpha}$ on spatially homogeneous time dependent test
functions. We presented a method for obtaining general expressions
for the action of the non local kernel on such test functions,
which only involves the zero mode solution of the Klein-Gordon
equation for a massive, minimally coupled scalar field. We
specialized the general expressions to particular cases, as test
functions defined as powers of the cosmic or conformal times. We
have also obtained an expression for the action of $\ln [-\Box]$
on time dependent test functions by considering the limit
$\alpha\rightarrow 0$.

The application of the operator $(-\Box)^{-\alpha}$ to functions
that depend on $t$ in Robertson Walker spacetimes, or functions
that depend on $r$ in static metrics with spherical symmetry, has
been considered previously by Hamber and Williams in
Refs.\cite{hw1,hw2}. Their approach was based on an analytic
continuation of the action of the derivative operator $(-\Box)^n$
for positive integers $n$ to the case $n\rightarrow -\alpha$. For
simple test functions in which the analytic continuation can be
carried out, this procedure gives the same answers than the non
local kernels defined in this paper. However, it is worth to note
that, within our approach based on the integral representation
(Eq.(\ref{integralrep})), it is in principle possible to evaluate
$(-\Box)^{-\alpha} f$ for {\it arbitrary} scalar functions in {\it
arbitrary} spacetimes. Moreover, this representation shows
explicitly the non local properties of the kernel and could also
be generalized to tensor functions.

In the cases of the FRW spacetimes in which both the scale factor
and the test function have a power law dependence on either the
cosmic or conformal time, after applying the non local kernels, we
have found that the functional dependence on time of the results
is the same as the one obtained after the replacement of the
energy scale $\bar{\mu}$ by the Hubble rate $H$. However, using as
an example the de Sitter spacetime, we have explicitly shown that
in general the result depend on the test function and can be very
different from the one obtained with the use of any of the
replacements mentioned in the Introduction.

We also investigated whether the action of $1/G(-\Box)$ on a
constant test function gives $1/G(0)$ or not, as naively expected
for analytic functions of the d'Alembertian. In particular, we
have shown that $(-\Box + b^2)^{-\alpha}1 = b^{-2\alpha}$ only for
spacetimes without initial singularities, or when considering
acausal propagators, and that it seems not to be related to the
existence or not of a particle horizon. These results are in tune
with the claim \cite{DGPAcaussal} that, in the presence of an
initial singularity, acausality may be crucial to solve the
cosmological constant problem using infrared modifications to
General Relativity.

As for future work, we consider that a similar method as the one
proposed in this paper could
also be used, without much additional difficulty, in static and
spherically symmetric spacetimes for test functions that respect
such symmetries.

\begin{acknowledgments}
We would like to thank G. Dvali for pointing Ref.\cite{Dvali06} to
us, and for helpful comments. We also thank G. Giribet for useful
discussions. This work has been supported by Universidad de Buenos
Aires, CONICET and ANPCyT.
\end{acknowledgments}

\appendix

\section{Expansion up to first order in $\alpha$ }\label{Appendix}

 In this appendix we
describe some technical details about the computation of the
 expansion up to first order in $\alpha$ of the application of $(-\Box)^{-\alpha}$
 to time dependent functions. The procedure we follow here is the same as the one
 described in the main text for the Minkowski case.

\subsection{De Sitter spacetime}
 In order to obtain the result presented
 in Eq.(\ref{LNdeSitter}), we start from Eq.(\ref{deSitterRes}) (in which the
zero mode Green's function $g^{\alpha}_b$ is given). We compute a
primitive of the  Green's function $g^{\alpha}_0$ with respect to
$\eta'$,
 \ba \mathcal{G}(\eta,\eta')&\equiv&\int d\eta'\;\mu^{2\alpha} g^{\alpha}_0 (\eta,\eta')=\frac{3^{1/2-\alpha}
 \mu^{2\alpha}}{H^{2\alpha}}\frac{\Gamma(1/2)}{\Gamma(\alpha)}\int^{\eta'}
 \frac{d\eta''}{-\eta''}\left(\frac{\eta}{\eta''}\right)^{3/2}
\frac{I_{\alpha-1/2}\left(\frac{3}{2}\ln\left(\frac{\eta''}{\eta}\right)\right)}{\left(\ln\left(\frac{\eta''}{\eta}\right)\right)^{1/2-\alpha}}
\\\nonumber &=&-\frac{2^{\alpha+1/2}\mu^{2\alpha}}{(3
H)^{2\alpha}}\frac{\Gamma(1/2)}{\Gamma(\alpha)}
\int^{3/2\ln(\eta'/\eta)} ds \; s^{\alpha-1/2}
I_{\alpha-1/2}(s)\exp\{-s\}\\\nonumber
&=&-\frac{3^{1/2-\alpha}\mu^{2\alpha}}{2(H)^{2\alpha}}\frac{\Gamma(1/2)}{\Gamma(\alpha+1)}\left(\frac{\eta}{\eta'}\right)^{3/2}\left(\ln\left(\frac{\eta'}{\eta}\right)\right)^{\alpha+1/2}\left\{I_{\alpha-1/2}\left(\frac{3}{2}\ln\left(\frac{\eta'}{\eta}\right)\right)+I_{\alpha+1/2}\left(\frac{3}{2}\ln\left(\frac{\eta'}{\eta}\right)\right)\right\},
 \ea where the second equality follows after the change of
 variable $s=3/2\ln(\eta''/\eta)$ (for the last equality see Ref.\cite{GR}).

Assuming that $\mathcal{G}(\eta,\eta')f(\eta')\sim
(\ln(\eta'/\eta))^{\alpha} f(\eta')$ goes to zero as
$\eta'\to-\infty$ ($\alpha\to 0^+$), and expanding
$\mathcal{G}(\eta,\eta')$ up to first order in $\alpha$, \be
\mathcal{G}(\eta,\eta')\simeq-1-\alpha\left\{\ln\left(\frac{e^{\gamma}\mu^2}{3
H^2}\ln\left(\frac{\eta'}{\eta}\right)\right)+E_i\left(-3\ln\left(\frac{\eta'}{\eta}\right)\right)\right\},\ee
we get \be \int d\eta'\; \mu^{2\alpha}g^{\alpha}_0 (\eta,\eta')
f(\eta')-f(\eta) \simeq \alpha \int_{-\infty}^{\eta} d\eta'
\left\{\ln\left(\frac{e^{\gamma}\mu^2}{3H^2}\ln\left(\frac{\eta'}{\eta}\right)\right)+E_i\left(-3\ln\left(\frac{\eta'}{\eta}\right)\right)\right\}
\frac{d}{d\eta'}f(\eta'), \ee where $E_i$ is the
exponential-integral function \cite{GR}. Then, the required
operator (\ref{LNdeSitter}) is obtaining by comparing this result
with the first order of the alpha expansion of
$(-\Box)^{-\alpha}$.

\subsection{FRW spacetime with $a(t)\varpropto t^{1/2}$}

We begin by computing a primitive of the Green's function
$g^{\alpha}_0$ given in Eq.(\ref{gRadiac}) (see Ref.\cite{GR}),
\ba \mathcal{G}(\eta,\eta')&\equiv&\int d\eta'\;
\mu^{2\alpha}g^{\alpha}_0 (\eta,\eta')=4^{1-\alpha}\frac{(\mu
t)^{2\alpha}}{\Gamma(2\alpha)}\int_0^{\eta'}\frac{{\eta''}^3}{\eta^4}
\left(1-\frac{{\eta''}^4}{\eta^4}\right)^{2\alpha-1}
F\left[\alpha-\frac{1}{4},\alpha,2\alpha;
1-\frac{{\eta''}^4}{\eta^4}\right]\\\nonumber
&=&\frac{\Gamma(5/4)\Gamma(\alpha+1/2) (\mu
t)^{2\alpha}}{\Gamma(\alpha+5/4)\Gamma(1/2)\Gamma(2\alpha+1)}-\frac{(\mu
t)^{2\alpha}}{4^{\alpha}\Gamma(2\alpha+1)}\left(1-\frac{{\eta'}^4}{\eta^4}\right)^{2\alpha}F\left[\alpha-\frac{1}{4},\alpha,2\alpha+1;
1-\frac{{\eta'}^4}{\eta^4}\right].\ea Then, the integration by
part results in \be \label{partint}\int
d\eta'\mu^{2\alpha}g^{\alpha}_0(\eta,\eta') f(\eta')=
\mathcal{G}(\eta,\eta)
f(\eta)-\int_{0}^{\eta}d\eta'\mathcal{G}(\eta,\eta')
\frac{d}{d\eta'}f(\eta'), \ee where we have assumed that
$f(\eta')\mathcal{G}(\eta,\eta')\sim f(\eta'){\eta'}^4\to 0$ as
$\eta'\to 0$. It is straightforward to calculate the expansion up
to first order in $\alpha$ of
$\mathcal{G}(\eta,\eta)$,\be\label{RGetaeta}\mathcal{G}(\eta,\eta)\simeq
1+\alpha(-4+2\gamma+\frac{\pi}{2}+\ln(2\mu^2 t^2)).\ee In order to
expand $\mathcal{G}(\eta,\eta')$ it is useful to use the following
identities (see Ref.\cite{Abram}): \ba
F\left[\alpha-\frac{1}{4},\alpha,2\alpha+1;
1-z\right]&=&\frac{\Gamma(2\alpha+1)}{\Gamma(\alpha+1)}\left\{\frac{\Gamma(5/4)}{\Gamma(\alpha+5/4)}F\left[\alpha-\frac{1}{4},\alpha,-\frac{1}{4};
z\right]\right.\\\nonumber &+&\alpha\;\left.
z^{5/4}\frac{\Gamma(-5/4)}{\Gamma(\alpha-1/4)}F\left[\alpha+\frac{5}{4},\alpha+1,\frac{1}{4};
z\right]\right\},\\
F\left[\alpha-\frac{1}{4},\alpha,-\frac{1}{4}; z\right]&=&1+\alpha
(1-4\alpha)\int_0^z
ds\;F\left[\alpha+\frac{3}{4},\alpha+1,\frac{3}{4};
s\right]\\\nonumber &\simeq& 1+\alpha\int_0^z
ds\;F\left[\frac{3}{4},1,\frac{3}{4}; s\right]=1-\alpha
\ln(1-z).\ea Then, we find that up to first order in $\alpha$
\be\label{RGetaetaprima} \mathcal{G}(\eta,\eta')\simeq
\alpha\left\{\frac{4{\eta'}^5}{5\eta^5}
F\left[\frac{5}{4},1,\frac{9}{4};\frac{{\eta'}^4}{\eta^4}\right]-\ln\left(1-\frac{{\eta'}^4}{\eta^4}\right)\right\}.
\ee Inserting (\ref{RGetaeta}) and (\ref{RGetaetaprima}) into
Eq.(\ref{partint}), and comparing with the expansion of
$(-\Box)^{-\alpha}$ we arrive at the result given in
Eq.(\ref{LNinRad}).

\subsection{FRW spacetime with $a(t)\varpropto t^{2}$}

A primitive of the Green's function given in Eq.(\ref{gtcuad}) can
be obtained as\ba \mathcal{G}(\eta,\eta')&\equiv&\int
d\eta'\;\mu^{2\alpha} g^{\alpha}_0 (\eta,\eta')=\int_0^t dt'
\frac{(t-t')^{2\alpha-1}\mu^{2\alpha}}{\Gamma(2\alpha)(1+2\alpha)(3+2\alpha)}\\\nonumber
&\times&\left[3\frac{ t'}{t}+3\frac{{t'}^5}{t^5}+3(2\alpha-1)
\left(\frac{{t'}^2}{t^2}+\frac{{t'}^4}{
t^4}\right)+(3+4\alpha(\alpha-1))\frac{{t'}^3}{t^3}\right], \ea
where we have used that $a(\eta')d\eta'= dt'$. In this case, it is
not difficult to carry out the $\alpha$ expansion to find that up
to first order we have
\ba \mathcal{G}(\eta,\eta)&\simeq& 1+\alpha\left\{-\frac{46}{15}+\ln(e^{2\gamma}\mu^2t^2)\right\},\\
\mathcal{G}(\eta,\eta')&\simeq& \alpha\left\{-\frac{2
t'}{t}-\frac{2{t'}^5}{5 t^5}-\frac{2{ t'}^3}{3
t^3}-2\ln\left(1-\frac{t'}{t}\right)\right\}, \ea where we have
assumed that $\mathcal{G}(\eta,\eta') f(t')\sim{} {t'}^2 f(t') \to
0$ as $t'\to 0$.

\end{document}